\documentclass[aps,twocolumn,PRB,reprint,superscriptaddress]{revtex4-1}

\usepackage{amsmath}
\usepackage{epsfig}
\usepackage{graphicx}
\usepackage{graphicx, color, epstopdf}
\usepackage{bm}
\usepackage{amssymb}
\usepackage{hyperref}
\usepackage{xcolor}
\usepackage{subfigure}
\usepackage{makecell}
\usepackage{tikz}

\usepackage{etoolbox}
\usepackage{graphicx}
\usepackage{amsfonts}
\usepackage{amsmath}
\usepackage{amsthm}
\usepackage{mathrsfs}
\usepackage{amssymb}
\usepackage{bm}
\usepackage{hyperref}
\usepackage{slashed}
\usepackage{framed}
\usepackage{xcolor}
\usepackage{extarrows}

\allowdisplaybreaks[4]

\hypersetup{hidelinks,
	colorlinks=true,
	allcolors=black,
	pdfstartview=Fit,
	breaklinks=true}
	
\definecolor{lime}{HTML}{A6CE39}
\DeclareRobustCommand{\orcidicon}{
\begin{tikzpicture}
\draw[lime, fill=lime] (0,0)
circle[radius=0.16]
node[white]{{\fontfamily{qag}\selectfont \tiny \.{I}D}};
\end{tikzpicture}
\hspace{-2mm}
}
\foreach \x in {A, ..., Z}{%
\expandafter\xdef\csname orcid\x\endcsname{\noexpand\href{https://orcid.org/\csname orcidauthor\x\endcsname}{\noexpand\orcidicon}}
}

\begin{document}
\bibliographystyle{apsrev4-1}
\widetext
\title{Diagnosing Altermagnetic Phases through Quantum Oscillations}
\author{Zhi-Xia Li}
\affiliation{National Laboratory of Solid State Microstructures, School of Physics,
Jiangsu Physical Science Research Center, and Collaborative Innovation Center of Advanced Microstructures,
Nanjing University, Nanjing 210093, China}
\author{Hanjing Zhou}
\affiliation{National Laboratory of Solid State Microstructures, School of Physics,
Jiangsu Physical Science Research Center, and Collaborative Innovation Center of Advanced Microstructures,
Nanjing University, Nanjing 210093, China}
\author{Xiangang Wan}
\affiliation{National Laboratory of Solid State Microstructures, School of Physics,
Jiangsu Physical Science Research Center, and Collaborative Innovation Center of Advanced Microstructures,
Nanjing University, Nanjing 210093, China}
\author{Wei Chen \hspace{-1.5mm}\orcidA{}}
\email{Corresponding author: pchenweis@gmail.com}
\affiliation{National Laboratory of Solid State Microstructures, School of Physics,
Jiangsu Physical Science Research Center, and Collaborative Innovation Center of Advanced Microstructures,
Nanjing University, Nanjing 210093, China}

\date{\today}
\begin{abstract}
The recently delimited altermagnetic phase is characterized by
zero net magnetization but momentum-dependent collinear spin-splitting.
To explore the intriguing physical effects and potential applications of altermagnets,
it is essential to analyze their Fermi surface properties, encompassing both configurations and spin textures.
Here, we conduct a Fermiology study on metallic altermagnets and demonstrate that
the collinear spin-split features of their
Fermi surfaces can be clearly revealed through quantum oscillation measurements.
By introducing a transverse Zeeman field to remove the spin-degenerate lines
in the momentum space, the Fermi surface
undergoes a Lifshitz transition, giving rise
to spin-flipped cyclotron motion between orbits with opposite spins.
Accordingly, the Onsager-Lifshitz quantization yields two sets of Landau levels,
leading to frequency splitting of the Shubnikov-de Haas oscillations in conductivity.
In the presence of spin-orbit coupling, the Zeeman field causes two separate cyclotron orbits to merge at the
Lifshitz transition point before splitting again. This results in the two original
frequencies discontinuously changing into a single frequency equal to their sum.
Our work unveils a unique and universal signature of altermagnetic Fermi surfaces
that can be probed through quantum oscillation measurements.

%The Fermi surface of a two-dimensions $d-$wave altermagnet is ellipse-shaped. In the presence of a in-plane Zeeman splitting, the Fermi surface becomes petal-shaped.  According to the semiclassical theories, if a perpendicular magnetic field is applied, electrons move along quantized semiclassical orbits and form discrete energy spectra, which are known as Landau levels. For a weak Zeeman splitting, there is a probability that electrons tunnel from one orbit of the Fermi surface to another, which is known as magnetic breakdown. We focus on the effective Hamiltonian near the breakdown regions and obtain the quantization conditions for the semiclassical orbits by considering the Landau-Zener tunneling problem. For a strong Zeeman splitting, the tunnel probability becomes zero and the quantization conditions give two sets of Landau levels. One can see two frequency components in the corresponding Shubnikov-de Haas oscillation, which is in contrast to the case without Zeeman splitting where only one frequency component can be seen. Such frequency transition of quantum oscillation induced by in-plane Zeeman splittings can be regarded as a fingerprint signature of altermagnets and can be detected experimentally.

\end{abstract}

\maketitle

\section{Introduction}
The investigation of magnetic systems has long been an active branch of condensed
matter physics. Among various magnetically ordered phases,
the collinear quantum magnets are usually divided into two phases,
ferromagnetism and antiferromagnetism~\cite{10.1126}.
Recently, a new type of collinear
magnetic phase dubbed altermagnetism has been proposed based on spin group theory and is
attracting increasing attention in condensed matter physics~\cite{PhysRevX.12.031042,PhysRevX.12.040501,PhysRevX.12.040002,foot}.
Distinct from ferromagnetism and conventional antiferromagnetism, such materials exhibit
large momentum-dependent spin splitting and zero net magnetization,
which stem from nonrelativistic spin and crystal rotation
symmetries~\cite{PhysRevX.12.031042,PhysRevX.12.040002,PhysRevX.12.040501,PhysRevX.12.011028}. The
spin-dependent Fermi surfaces of altermagnets may exhibit $d$-, $g$-, or $i$-wave symmetries,
which lead to many intriguing physical properties, such as spin current~\cite{Shao2021,PhysRevLett.126.127701,Bose2022},
anomalous Hall effect~\cite{vsmejkal2020crystal,feng2022anomalous,guo2023quantum,Hayami21prb}, crystal magneto-optical Kerr effect~\cite{PhysRevB.92.144426}
and giant magnetoresistance \cite{PhysRevX.12.011028} that cannot occur
in conventional antiferromagnets.
Altermagnets are predicted to
have various potential applications, including
spintronics~\cite{PhysRevLett.130.216701,PhysRevLett.128.197202,PhysRevLett.129.137201},
correlated states of matter~\cite{PhysRevX.12.040501}, superconductivity~\cite{PhysRevLett.131.076003,PhysRevB.108.054511}, \emph{etc}..

\begin{figure}[h]
	\centering
	\includegraphics[width=0.45\textwidth]{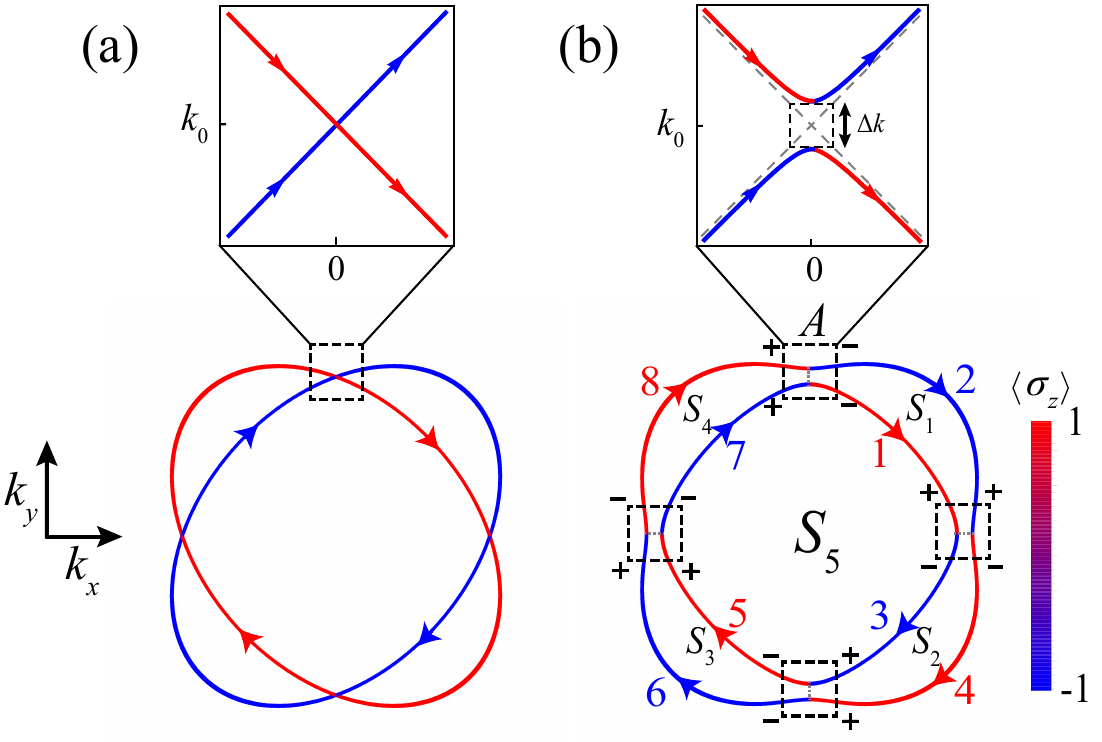}
	\caption{Schematic illustration of cyclotron motion along Fermi surfaces (a) {without} and (b) {with}
 the in-plane Zeeman field. The upper panels are zoom-in images at the transition regions
 denoted by the dashed boxes. The average value $\langle\sigma_z\rangle$
 is measured by the color bar. The numbers in (b) label the line segments
 and $S_{1,\cdots,5}$ denote the areas of respective regions encircled by them.
 The incoming and outgoing states are labeled by $+$ and $-$, respectively.}
\label{fig1}
\end{figure}

Hundreds of material candidates have been predicted to be altermagnetic~\cite{Savitsky},
including three-dimensional compounds such as MnTe, $\rm{RuO_2}$, CrSb, and $\rm{La_2CuO_4}$~\cite{PhysRevX.12.040501,PhysRevB.107.L100418,PhysRevB.107.L100417,PhysRevLett.131.256703}, and
two-dimensional monolayer $\rm{RuF_4}$ with the $d$-wave altermagnetic symmetry~\cite{Milivojevic2024}.
Very recently, the altermagnetic lifting of Kramers spin degeneracy has been
observed in $\rm{RuO_2}$ and MnTe (MnTe$_2$) by angle-resolved photoemission spectroscopy
(ARPES) or spin-ARPES~\cite{PhysRevB.109.115102,2024arXiv240204995L,krempasky2024altermagnetic,zhu2024observation,yang2024three,ding2024large,zeng2024observation},
demonstrating the $d$-wave spin pattern in $\rm{RuO_2}$~\cite{2024arXiv240204995L} and $g$-wave
spin pattern in CrSb~\cite{yang2024three,ding2024large,zeng2024observation}.
Meanwhile, the muon spin rotation and relaxation experiment on $\rm{RuO_2}$ has
reported the absence of magnetic order~\cite{PhysRevLett.132.166702,kessler2024absence}, casting doubt on
the diagnosis of altermagnetic phase.
Consequently, further measurements such as quantum oscillations, are necessary to ascertain the nature of altermagnets.
Moreover, to understand the intriguing physical effects of the altermagnets and explore
their potential applications, it is essential to analyze the geometric
and spin configurations of altermagnetic Fermi surfaces.

In this work, we study quantum oscillations in metallic altermagnets
to uncover the distinctive properties of their Fermi surfaces. By imposing
an in-plane Zeeman field to remove spin degeneracy along specific lines in
the momentum space which then causes a Lifshitz transition of the Fermi surface~\cite{lifshitz1960anomalies},
the altermagnets can be diagnosed through the frequency transitions of Shubnikov-de Haas oscillations
in conductivity. Specifically, as the Zeeman field increases, the cyclotron
orbits change from two identical orbits for both spin polarizations to two
reconstructed ones with spin hybridization. As a result, the Onsager-Lifshitz
quantization~\cite{108014786440908521019,lifshitz1956theory} along these cyclotron
orbits results in splitting of Landau levels,
which is manifested as the Zeeman-field induced frequency splitting of
the Shubnikov-de Haas oscillations.
When spin-orbit coupling (SOC) exists,
a $\pi$ Berry phase is accumulated along two pristine cyclotron orbits.
In such a regime, a Zeeman field can drive two
semiclassical orbits to merge together,
leading to an discontinuously change of the oscillation frequency from two components
to a single one. Our results can be generally applied to
various material candidates of altermagnets, thus providing
an effective and universal approach for its identification.

\section{Spin-flipped cyclotron motion}
To be specific, we consider a 2D $d$-wave metallic altermagnet
with a Zeeman field along the $x$-direction, which can be described by the Hamiltonian as
\begin{equation}
H(\bm{k})=a(k_x^2+k_y^2)+Jk_xk_y\sigma_z+\Delta\sigma_x,
\label{eq1}
\end{equation}
where $\boldsymbol{k}=(k_x, k_y)$ is the momentum, $\sigma_{x,y,z}$ are the spin Pauli matrices,
$a$ and $J$ parameterize the kinetic term and the $d$-wave exchange interaction, respectively.
$\Delta=\mu_B B_x g^*/2$ is the Zeeman energy splitting~\cite{Fernandes24prb} with $B_x$ the Zeeman
field and $g^*$ the Land\'{e} $g$ factor.
The corresponding eigenenergies are
$
E_{\pm}(k)=a(k_x^2+k_y^2)\pm \sqrt{J^2k_x^2k_y^2+\Delta^2}.
$
In the absence of a Zeeman field ($\Delta=0$), the Fermi surface is composed of two intersecting
ellipses corresponding to two opposite spins [Fig.~\ref{fig1}(a)]; A finite Zeeman field hybridizes the two spin bands
and accordingly, the Fermi surface undergos a Lifshitz transition characterized by a momentum separation $\Delta k$
at the originally degenerate points [Fig.~\ref{fig1}(b)].

As a perpendicular magnetic field $B$ in the $z$ direction is applied, electrons are driven to
move along equienergy contours (labeled by $O$). In the absence of tunneling
between different orbits, the eigenenergies are selected by the following
quantization condition~\cite{108014786440908521019,lifshitz1956theory}
\begin{equation}
S(O)=2\pi l^{-2}(n+\gamma),\quad n=0,1,2...,
\label{eq3}
\end{equation}
where $S$ is the area encircled by the cyclotron orbits in the momentum space,
$l=\sqrt{\hbar/eB}$ is the magnetic length, and $\gamma=1/2$ is the Maslov index~\cite{PhysRevB.97.144422}.

More generally, a magnetic breakdown may occur between the outer and inner Fermi contours
in Fig.~\ref{fig1}(b)~\cite{Pippard1969}, which leads to more complicated
quantization rules compared with Eq.~\eqref{eq3}. To analyze the magnetic breakdown,
we adopt the Landau gauge for the vector potential as $\boldsymbol{A}=(0,Bx,0)$ and make the substitution
$k_y\rightarrow k_y+Bx$ in Eq.~\eqref{eq1}, where $\hbar=1, e=1$ is set for simplicity.
In the momentum representation, the coordinate operator reads $x\rightarrow i\partial_{k_x}$.
Quantum tunneling may occur in the breakdown regions marked with black dashed squares in Fig.~\ref{fig1}(b).
Given the symmetry of the Fermi surface, it is sufficient to analyze one specific region A
and then combine the results of four regions. In region A,
$k_x\simeq0$, $k_y\simeq k_0=\sqrt{E_F/a}$ with $E_F$ the Fermi energy, and the Hamiltonian~\eqref{eq1}
can be expanded at $(k_x, k_y)=(0, k_0)$ to the linear order of momentum as
\begin{equation}
\mathcal{{H}}(\bm{k})=ak_0(2k_y-k_0)+ Jk_0k_x \sigma_z+\Delta\sigma_x,
\label{eq4}
\end{equation}
which gives two hyperbola segments of Fermi surface as shown in the upper panel of
Fig.~\ref{fig1}(b). As shown in Appendix \ref{A}, by interpreting $k_x\equiv t$ as time,
the problem of magnetic breakdown can be mapped to the model of Landau-Zener
tunneling. A direct calculation yields the tunnel probability $Z=\text{exp}(-\frac{\pi}{B}\frac{\Delta^2}{2aJk_0^2})$~\cite{SHEVCHENKO20101},
which will be used to determine the quantization conditions.

\section{Quantization conditions and Landau levels}
We regard $k_y$ as a parameter
and solve the state evolution in region A in the $k_x\equiv t$ representation.
Consider an electron entering region A from $k_x<0$
and exiting to $k_x>0$; see the upper panel of Fig.~\ref{fig1}(b).
A straightforward derivation in Appendix \ref{B} yields the scattering matrix $\mathcal{S}$
that connects the incident and outgoing waves as~\cite{040721-021331}
\begin{equation}
\begin{aligned}
\left(\begin{array}{cccccc}
c_{2}^- \\
c_{1}^-\\
\end{array}\right)
= \mathcal{S}\left(\begin{array}{cccccc}
c_{8}^+\\
c_{7}^+\\
\end{array}\right),
\mathcal{S}=\left(\begin{array}{cccccc}
\sqrt{1-Z}e^{-i\omega}  & -\sqrt{Z} \\
\sqrt{Z} & \sqrt{1-Z}e^{i\omega}\\
\end{array}\right),
\end{aligned}
\label{eq5}
\end{equation}
where the amplitudes $(c_8^+, c_7^+)^{\text{T}}$ and $(c_2^-, c_1^-)^{\text{T}}$ correspond to the incoming ($+$)
and outgoing ($-$) states, respectively [cf. Fig.~\ref{fig1}(b)].
$\omega$ is the so-called Stokes phase~\cite{Wubs_2005} which reads
$
\omega={\pi}/{4}+\delta(\ln\delta-1)+\text{Arg}[\Gamma(1-i\delta)]$,
with
$
\quad \delta={\Delta^2}/{(4BaJk_0^2)}
$
and $\Gamma(\cdot)$ the gamma function.

\par
A full cyclotron motion constitutes four such scattering processes connected by free propagations
along the line segments denoted by $i=1,\cdots,8$ in Fig.~\ref{fig1}(b),
in which the areas surrounded by the line segments are denoted by $S_{1,\cdots,5}$.
Due to the $C_4$ rotation symmetry of the orbits, the four equal areas are
set to $S_{1,\cdots,4}=S_0$. Define $\Theta_i$ as the phase accumulation
during the free propagation along the line segment $i$. Since the amplitudes $\{c_i^+, c_i^-\}$
are everywhere single-valued, we arrive at the following equation
\begin{equation}
	\begin{aligned}
	\text{det}\left\{\prod_{j=1}^4\left[\mathcal{S}\left(\begin{array}{cccccc}
	e^{i\Theta_{2j}} & 0\\
	0 & e^{i\Theta_{2j-1}}\\
	\end{array}\right)\right]
	-\mathcal{I}
	\right\}=0,
	\end{aligned}
	\label{eq8}
	\end{equation}
where the matrix product is arranged from right to left as $j$ increases,
and $\mathcal{I}$ is the identity matrix.
For any cyclotron orbit $O$, the accumulated
phase during free propagation is given by~\cite{PhysRevB.97.144422}
$
\sum_{i\in O}\Theta_i=l^2S(O)-2\pi\gamma,
$
where the sum is taken over the line segments that constitute the orbit $O$.
As shown in Appendix \ref{B}, solving Eq.~\eqref{eq8} yields the quantization condition
\begin{equation}
\begin{aligned}
&\cos[l^2(S_5+2S_0)-2\pi\gamma]-(1-Z)^2\cos(2l^2S_0-4\omega)\\
&\ \ \ +4Z(1-Z)\cos(l^2S_0-2\omega)+Z(2-3Z)=0.
\end{aligned}
\label{eq10}
\end{equation}
For an arbitrary $Z$, the quantized energy spectra can be obtained by inserting the explicit functions
$S_5=S_5(E)$ and $S_0=S_0(E)$ of energy $E$ into Eq.~\eqref{eq10}.

\par
Our main focus is on the two semiclassical limits $Z\rightarrow1$ (magnetic breakdown regime) and $Z\rightarrow0$ (adiabatic regime),
which correspond to scenarios with small and large Zeeman fields, respectively.
In these
two limits, the quantization conditions reduce to
\begin{equation}
\begin{aligned}
Z\rightarrow1: S_5+2S_0&=2\pi l^{-2}(n+\gamma),\\
Z\rightarrow0: S_5+4S_0&=2\pi l^{-2}(n+\gamma),\
{\rm or}\ \  S_5=2\pi l^{-2}(n+\gamma),\\
\end{aligned}
\label{eq11}
\end{equation}
where we have considered in the second line that $\omega\rightarrow0$ as $Z\rightarrow0$.
The quantization condition for $Z\rightarrow1$ correspond to the two identical ellipse-shaped orbits in
Fig.~\ref{fig1}(a), which yields the spin-degenarate Landau levels $E_n=(eB/\hbar)\sqrt{4a^2-J^2}(n+\gamma)$.
In the $Z\rightarrow0$ limit, the two quantization conditions correspond to the outer and inner orbits respectively
in Fig.~\ref{fig1}(b), which give rise to two different sets of Landau levels.

To verify such observations, we numerically calculate the energy levels as a function of $B$
in the two limits. To this end, we interpret $x$ and $k_x$
by the ladder operators as $x=\sqrt{\hbar/2eB}(\pi+\pi^{\dag})$ and $k_x=i\sqrt{eB/2\hbar}(\pi^\dag-\pi)$,
which operate on the Landau level basis through $\pi|n\rangle=\sqrt{n}|n-1\rangle$ and
$\pi^{\dag}|n\rangle=\sqrt{n+1}|n+1\rangle$. A proper cutoff is adopted for $n$ in the calculation
to ensure the convergence of the results.
In the rest of this paper, $k_x$ and $x$ are set to be dimensionless and accordingly,
$a$ and $J$ have the units of energy.
The Landau levels in both limits, as a function of $B$, are shown in Fig. \ref{fig2}(a-b).
Specifically, there is only one set of Landau levels with spin degeneracy in the limit $Z\rightarrow1$ as shown Fig.~\ref{fig2}(a);
In the opposite limit $Z\rightarrow0$, two sets of Landau levels become visible in Fig.~\ref{fig2}(b),
corresponding to the quantized outer and inner orbits in Fig.~\ref{fig1}(b).
Away from the two semiclassical limits, the typical energy spectra
solved by the quantization condition~\eqref{eq10} exhibit a quasirandom feature~\cite{PhysRevB.97.144422}.
Recall that an electron undergoes four tunneling events along the $Z\rightarrow1$ orbit and none along the $Z\rightarrow0$ orbit. For the electron to form a closed orbit, the number of tunnelling events must be even. Therefore, for intermediate values of $Z$, the electron can also undergo two tunneling events, corresponding to other two types of orbits with enclosed areas $S_5+S_0$ and $S_5+3S_0$, respectively.

\section{Shubnikov-de Haas oscillations}
The analysis
of the Landau levels indicates that the main features
of altermagnetic Fermi surface can be well reflected by
the evolution of the quantum oscillations as the Zeeman field varies.
Here, we focus on the Shubnikov-de Haas oscillations in conductivity.
As $B$ changes, whenever the
Fermi level is aligned with one of the Landau levels, the density of states (DOS)
reaches a maximum, thereby leading to an oscillating behavior in resistivity.
Using the linear response theory, the conductivity tensor reads~\cite{PhysRevLett.117.077201,quantum}
\begin{equation}\label{eqs48}
\sigma_{\mu\nu}=\frac{e^2\hbar}{\pi V}\int dE f(E) \text{Re}\left\{\text{Tr}\left[\hat{v}_\mu \frac{\partial \hat{G}^{R}_{E}}{\partial E}  \hat{v}_\nu(\hat{G}^A_E-\hat{G}^R_E)\right]\right\},
\end{equation}
where the subscripts $\mu(\nu)=x,y$ and accordingly, $\sigma_{xx}$ and $\sigma_{xy}$ represent the longitudinal
and Hall conductivity, respectively. $f(E)=\left[e^{(E-E_F)/kT}+1\right]^{-1}$ is the Fermi-Dirac
distribution function, $\hat{v}_{\mu}=\partial_{k_{\mu}}H(\bm{k}+\bm{A})$ is the velocity operator,
and $\hat{G}^{R,A}=\left[E-H(\bm{k}+\bm{A})\pm i\chi\right]^{-1}$ are the retarded (+) and advanced ($-$) Green operators
with $\chi$ the level broadening.
Eq.~\eqref{eqs48} is evaluated by inserting the complete Landau level basis $\{|n\rangle\}$
and expressing the velocity operator by the ladder operators $\pi,\pi^\dag$.
The corresponding longitudinal resistivity
is given by $\rho_{xx}=\sigma_{xx}/(\sigma^2_{xx}+\sigma^2_{xy})$.
The reduced magnetoresistivity is defined as the difference between the resistivity in the presence
of a magnetic field and the resistivity without a magnetic field, normalized by the latter.
Specifically, $\delta \rho_{xx}(B)=[\rho_{xx}(B)-\rho_{xx}(0)]/\rho_{xx}(0)$.

\begin{figure}[h]
	\centering
	\includegraphics[width=1\columnwidth]{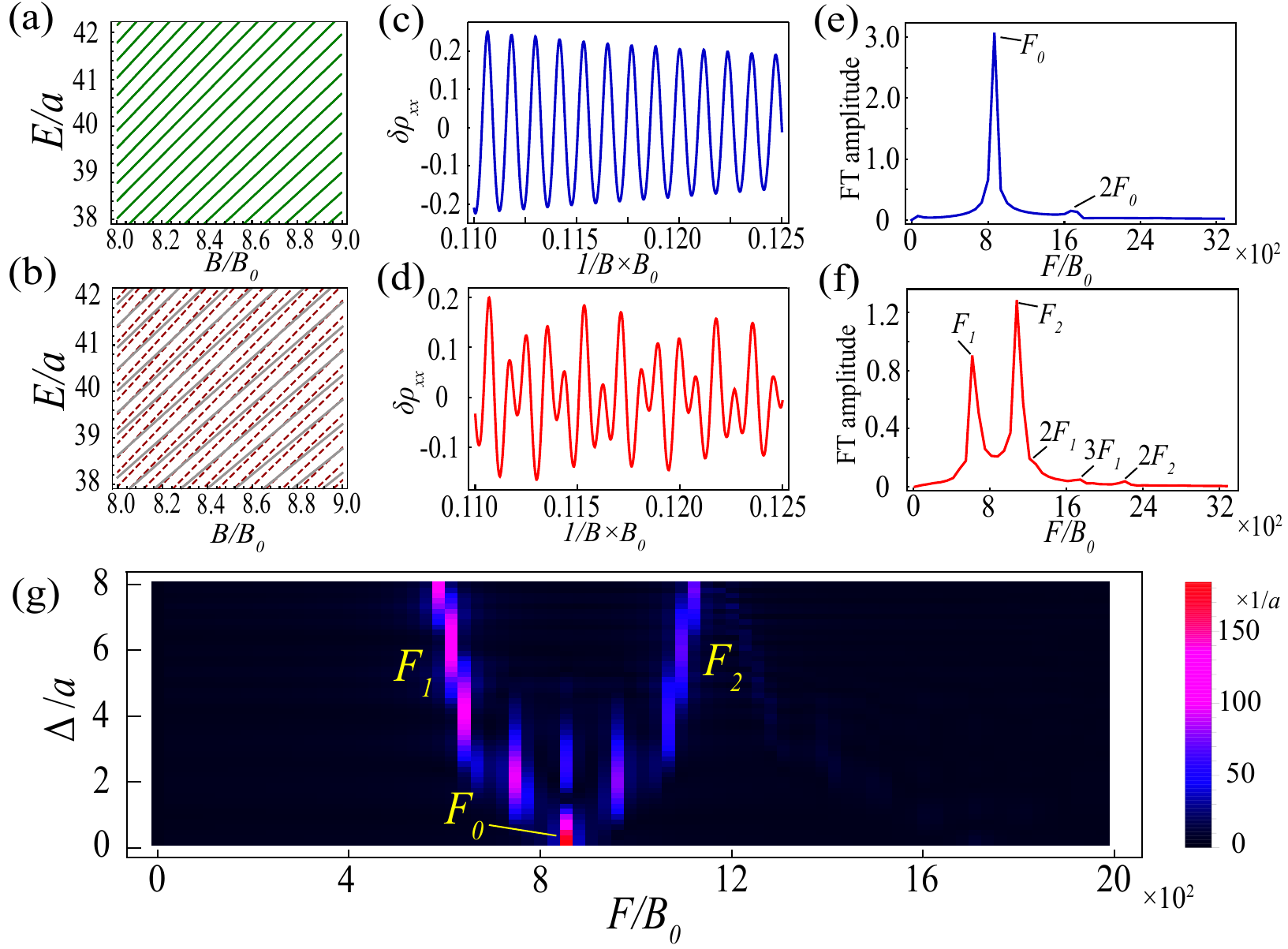}
	\caption{(a-b) Landau levels as a function of $B$ in (a) the $Z\rightarrow1$ limit with $\Delta=0$
and $Z\rightarrow0$ limit with $\Delta=5.6a$.
(c-f) The normalized longitudinal magnetoresistivity $\delta\rho_{xx}$ as a function of $1/B$ in the limit
 of (c) $Z\rightarrow1$ ($\Delta=0$) and (d) $Z\rightarrow0$ ($\Delta=5.6a$). (e) and (f) are
 The Fourier transform (FT) components of (c) and (d), respectively.
 (g) Fourier components of the oscillating DOS $\delta D(E_F,B)$ as a function of $\Delta$.
 The parameters are $J=0.7a$, $E_F=40a$, $B_0=\hbar a/eE_F$, $\Sigma=0.2a$, $\chi=0.16a$, $kT=0.04a$, and $e\mathcal{A}/\hbar\pi^{3/2}=1$.}
\label{fig2}
\end{figure}

The numerical results for $\delta\rho_{xx}$ as a function of $1/B$ in the two limits $Z\rightarrow1$
and $Z\rightarrow0$ are shown in Fig.~\ref{fig2}(c) and Fig.~\ref{fig2}(d), respectively,
with Fig.~\ref{fig2}(e) and Fig.~\ref{fig2}(f) being their Fourier components.
One can see that for $Z\rightarrow1$, the oscillation of $\delta\rho_{xx}$
consists of one main frequency component $F_0$ and its multiples with much weaker amplitudes,
corresponding to the contribution from only one set of Landau levels in Fig.~\ref{fig2}(a).
For $Z\rightarrow0$, the oscillation contains two main frequency components
$F_1$ and $F_2$, as well as their multiples, corresponding to the two sets of
Landau levels in Fig. \ref{fig2}(b), which is induced by the Zeeman field.
To reveal the frequency transition induced by the Zeeman field,
it is sufficient to calculate the DOS at the Fermi level $E_F$.
This can be approximated using a Gaussian function with a finite broadening ($\Sigma$) as
$
D(E_F,B)=\frac{eB\mathcal{A}}{\hbar\pi}\sum_{n}\sqrt{\frac{1}{\pi\Sigma^2}}{\rm exp}\left[-\frac{(E_F-E_n)^2}{\Sigma^2}\right],
$
where the prefactor is the degeneracy of Landau levels,
$\mathcal{A}$ is the area of the system, and $n$ is the Landau level index.
The Fourier component of its oscillating part, $\delta D(E_F,B)=D(E_F,B)-D(E_F,0)$, as a function of $\Delta$
is shown in Fig.~\ref{fig2}(g). One can see that only one frequency $F_0$ exists
for $\Delta\sim0$. As $\Delta$ increases, the frequency pattern evolves and finally
splits into two branches $F_1$ and $F_2$. For intermediate values of $\Delta$, one can also notice the frequencies corresponding to the orbits with areas $S_5+S_0$ and $S_5+3S_0$.
Such a frequency transition effectively
manifests the collinear spin-splitting of the altermagnetic Fermi surfaces
and so can serve as its fingerprint signature.

\par

\section{The effect of spin-orbit coupling}
\begin{figure}[h]
\centering
\includegraphics[width=8.6cm]{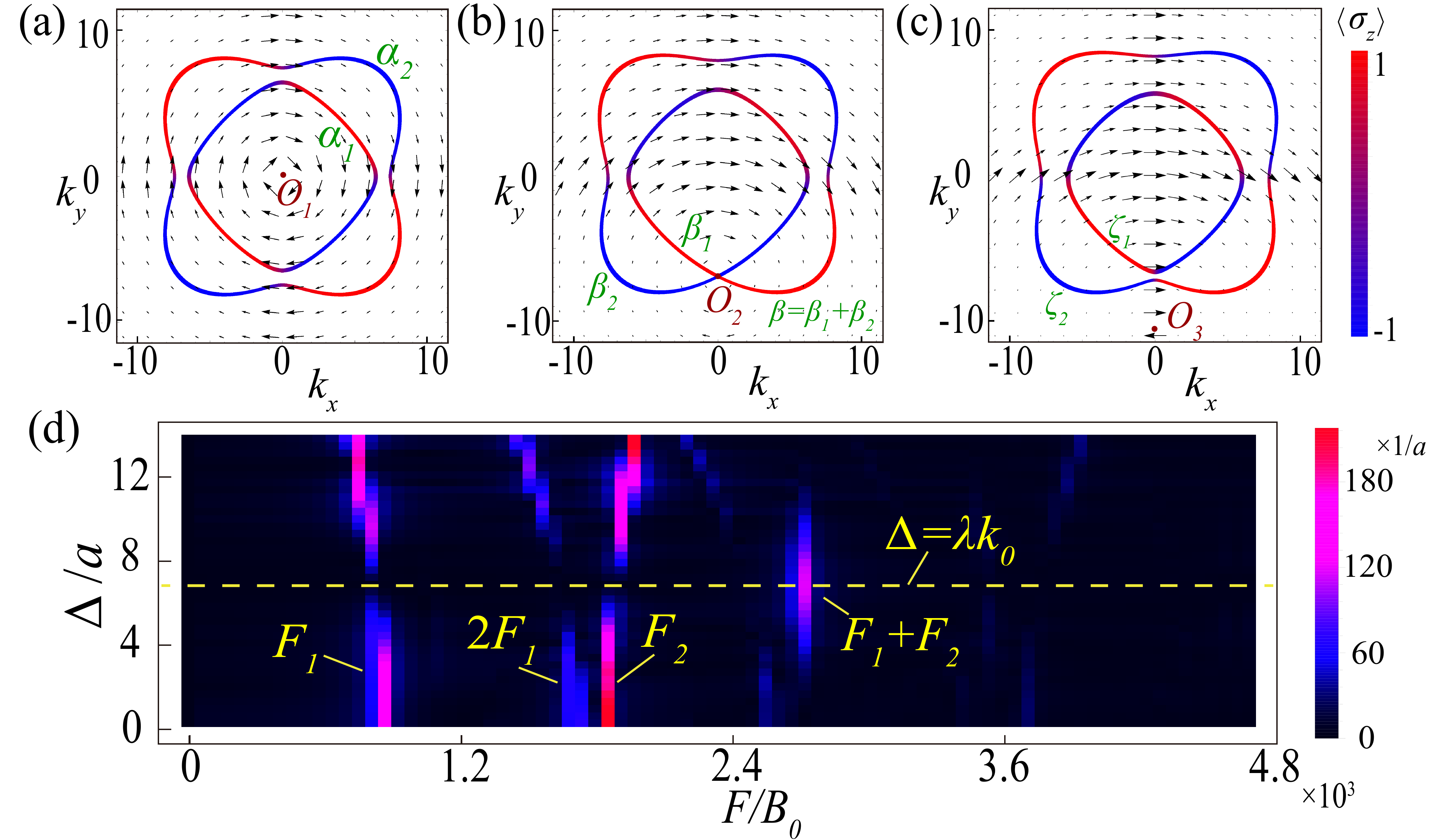}
\caption{(a-c) Evolution of Fermi surfaces and upper-band spin textures of
altermagnets with SOC
as the Zeeman field increases from (a) $\Delta=0$ to (b) $\Delta=\lambda k_0=\sqrt{48}a$
and finally to (c) $\Delta=11a$. The in-plane spin textures are denoted by the arrows
while its $z$ component is represented by the color.
$O_{1,2,3}$ are the centers of the spin vortices during the evolution and
$\alpha_{1,2}, \beta_{1,2}, \beta, \zeta_{1,2}$ label the cyclotron orbits.
(d) Fourier amplitudes of the oscillating DOS $\delta D(E_F,B)$ in the presence of
SOC. The critical value for the Zeeman field $\Delta=\lambda k_0$
is marked with the yellow dashed line. The parameters are $J=a$, $\lambda=a$
and $E_F=a k_0^2=48a$, $B_0=\hbar a/eE_F$. The Landau level broadening $\Sigma$
in (d) is set to $0.12a$.}
\label{fig3}
\end{figure}

In certain material candidates of altermagnets,
the SOC effect cannot be neglected, which can be
described by the whole Hamiltonian $H(\bm{k})+H_R(\bm{k})$ with
$
H_R(\bm{k})=\lambda(k_y\sigma_x-k_x\sigma_y)
$
the Rashba SOC of strength $\lambda$~\cite{PhysRevX.12.040501}. {Here, the Zeeman
field is applied along the $x$ axis. The eigenenergies are $\mathcal{E}_{\pm}=a(k_x^2+k_y^2)\pm\sqrt{(\Delta+\lambda k_y)^2+(\lambda k_x)^2+(Jk_xk_y)^2}$.}
In the absence of in-plane Zeeman term ($\Delta=0$), the SOC can still induce
a petal-shaped Fermi surface similar to that in Fig.~\ref{fig1}(b),
but with a vortical spin texture, as shown in Fig.~\ref{fig3}(a).
The vortical spin winding indicates that as an electron
moves along the outer or inner cyclotron orbit,
it will feel a nontrivial Berry phase $\phi_{B}=\mp\pi$ as shown in Appendix \ref{C}.
Consider the Rashba SOC is strong enough so that
the magnetic breakdown can be neglected, the
semiclassical quantization of the two orbits $\alpha_1$ and $\alpha_2$ in Fig.~\ref{fig3}(a) is
\begin{equation}
\begin{aligned}
S(\alpha_{1,2})=2\pi l^{-2}(n+\gamma-\phi_{B}/2\pi).
\end{aligned}
\label{eq16}
\end{equation}

As the Zeeman field $\Delta$ increases from zero, the center of the spin vortex
shifts downward from the origin, following the path $O_1\rightarrow O_2\rightarrow O_3$ in Figs.~\ref{fig3}(a-c),
along with the gap closing and reopening at the point  $O_2=(0, -k_0)$.
In particular, at the Lifshitz transition point where $\Delta=\lambda k_0$,
the two Fermi surfaces converge at $O_2$, coinciding with
the center of the spin vortex. Meanwhile, the tunnel probability between the outer and inner orbits at the
$O_2$ saturates with $Z=1$, indicating that the two
orbits, $\beta_1$ and $\beta_2$, merge into a single orbit, $\beta=\beta_1+\beta_2$, as shown in Fig.~\ref{fig3}(b). In such a non-adiabatic regime, the geometric phase can be
obtained by analyzing its asymptotic behavior on both sides.
Consider the asymptotic limit $\Delta=\lambda k_0-0^+$, both orbits $\beta_{1,2}$
encircle the spin vortex, resulting in the accumulation of a $(\mp)\pi$ phase. However,
the accumulated geometric phase along the entire $\beta$ orbit is zero,
as the phases along $\beta_1$ and $\beta_2$ cancel each other out.
In the opposite asymptotic limit $\Delta=\lambda k_0+0^+$,
both $\beta_{1,2}$ orbits have zero geometric phase, leading to a total phase of zero.
We conclude that at the critical point $\Delta=\lambda k_0$, no geometric phase
accumulates along the merged orbit $\beta$, resulting in the quantization condition
\begin{equation}
\begin{aligned}
S(\beta)=S(\beta_1)+S(\beta_2)=2\pi l^{-2}(n+\gamma).
\end{aligned}
\label{eq17}
\end{equation}

When $\Delta>\lambda k_0$, the two orbits separate again, and the center of the spin vortex
shifts outside of both Fermi contours, as shown in Fig.~\ref{fig3}(c).
As a result, the Berry phase accumulated along the two semiclassical orbits
$\zeta_1$ and $\zeta_2$ becomes zero, with the corresponding quantization condition being
$
S(\zeta_{1,2})=2\pi l^{-2}(n+\gamma).
$

\begin{figure*}[h]
\centering
\includegraphics[width=18cm]{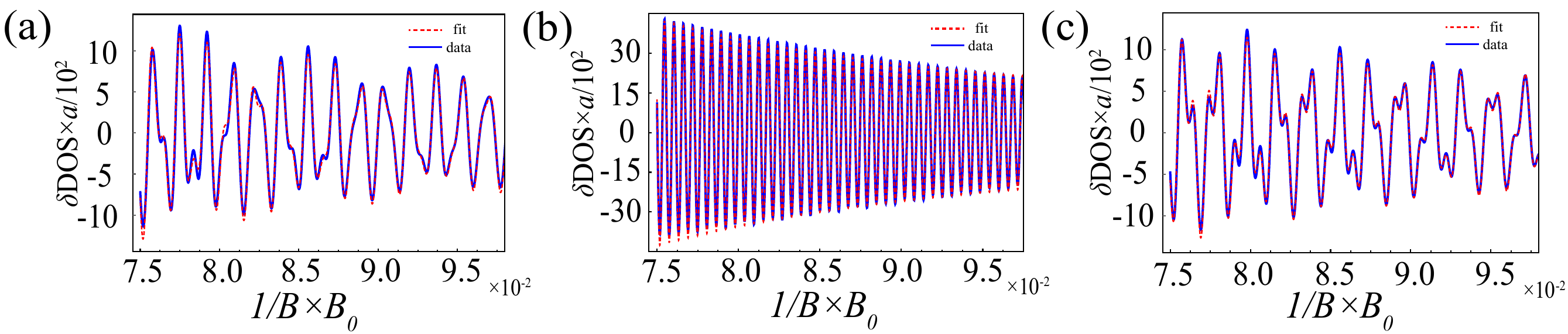}
\caption{Non-linear fittings (red dashed lines) of $\delta D(E_F,B)$ (solid blue lines) for (a) $\Delta=0$, (b) $\Delta=\lambda k_0=\sqrt{40}a$ and (c) $\Delta=12a$. $J=0.7a$, $\lambda=a$, $E_F=40a$, $\Sigma$ is set to $\Sigma=0.2a$ in (a,c) and $\Sigma=0.1a$ in (b) to achieve a visible oscillation in the latter with a smaller Landau level spacing.}
\label{figs1}
\end{figure*}

We verify the above analysis again by numerically
calculating the oscillation of the
DOS, $D(E_F,B)$, in the presence of SOC.
The corresponding Fourier amplitudes are shown in
Fig.~\ref{fig3}(d). As one can see, for $\Delta\ll\lambda k_0$
and $\Delta\gg\lambda k_0$, there exist two main
frequencies $F_1$ and $F_2$
steming from the contributions from two separate semiclassical orbits,
along with their multiples.
By contrast, for $\Delta\simeq\lambda k_0$,
$F_1$ and $F_2$ disappear and instead, a visible
frequency $F_1+F_2$ shows up, corresponding to the
combined orbit $\beta$ in Fig.~\ref{fig3}(b) and the quantization condition~\eqref{eq17}.
We conclude that in the presence of the SOC effect, {Zeeman fields along specific directions can induce frequency
transition with an enriched structure},
which is another distinctive signature of altermagnets.
Additionally, the associated $\pi$ phase shift across the Lifshitz transition point can also
be extracted from the oscillation patterns.
To extract the phase shift, we conduct a non-linear fitting for $\delta D(E_F,B)=D(E_F,B)-D(E_F,0)$
using the following fitting function~\cite{PhysRevLett.117.077201}
\begin{equation}
\begin{aligned}
\delta D_{\text{fit}}(E_F,B)=C_1e^{-N_1/B}\cos[2\pi(F_1/B+\phi_1)]\\
+C_2e^{-N_2/B}\cos[2\pi(F_2/B+\phi_2)]
\end{aligned}
\end{equation}
where the frequencies $F_1$ and $F_2$ are calculated by the Onsager-Lifshitz
relation $F_{1,2}=(\hbar/2\pi e)S_{1,2}$, with $S_{1,2}$
the areas of the two Fermi surfaces, and $C_1, C_2, N_1, N_2, \phi_1, \phi_2$ are the
fitting parameters.
The results of $\delta D_{\text{fit}}(E_F,B)$ for different Zeeman field are shown by the red dashed lines in Fig.~\ref{figs1}.
Specifically, the fitting functions in Figs.~\ref{figs1}(a-c) are
\begin{equation}\label{eqs59}
\begin{aligned}
\delta D(E_F,B)&=u_1e^{-w_1/B}\cos[2\pi(F_1/B-0.04534)]\\
+u_2e^{-w_2/B}&\cos[2\pi(F_2/B+0.04136)],\ \ (\Delta=0),\\
\delta D(E_F,B)&=u_3e^{-w_3/B}\cos[2\pi(F_3/B+0.5001)],\\
& (\Delta=\lambda k_0), \\
\delta D(E_F,B)&=u_4e^{-w_4/B}\cos[2\pi(F_4/B+0.4808)]\\
+u_5e^{-w_5/B}&\cos[2\pi(F_5/B+0.5173)],\ \ (\Delta>\lambda k_0),
\end{aligned}
\end{equation}
where $u_{1,2,3,4,5}=148.81$, 5.1974, 141.94, 100.10, 12.679, $w_{1,2,3,4,5}=5.0843$, 13.869, 9.0081, 6.2608, 11.329, $F_{1,2,3,4,5}=97.341$, 174.31, 271.65, 81.759, 189.89. Within a reasonable range of accuracy, one can find a phase shift $(\phi_{1,2}-\gamma)/(2\pi)\sim1/2$ compared to the Maslov index $\gamma=1/2$ for Fig.~\ref{figs1}(a), indicating a $\pi$ Berry phase. By contrast, in Fig.~\ref{figs1}(b) and Fig.~\ref{figs1}(c), the phase shift is zero, \emph{i.e.}, $(\phi_{1,2}-\gamma)/(2\pi)\sim0$. In experiments, such a phase shift can be extracted from the oscillation patterns.

{It should be noted that the above analysis holds true only when the Zeeman field is applied in the specific $x$ or $y$ direction. If the Zeeman field is applied along a general direction, with nonzero components in both directions, \emph{i.e.}, $\Delta_x\neq0$ and $\Delta_y\neq0$, the Lifshitz transition does not occur because the gap, given by $\sqrt{(\Delta_x+\lambda k_y)^2+(\Delta_y+\lambda k_x)^2+(Jk_xk_y)^2}$, remains nonzero. Therefore, by rotating the direction of the Zeeman field in the $x$-$y$ plane over a full cycle and varying its magnitude, the frequency combination will alternately appear and disappear four times. This provides a more distinct signature for the altermagnets.}

\section{DISCUSSION AND CONCLUSIONS}
The experimental identification of
the altermagnetic phase remains controversial at present~\cite{PhysRevB.109.115102,2024arXiv240204995L,
PhysRevLett.132.166702,kessler2024absence}. Therefore, additional evidence from
various approaches is crucial for its diagnosis. Quantum oscillation measurements is a widely used
method for characterizing the properties of the Fermi surface,
including its configuration and band topology, \emph{etc.}~\cite{040721-021331}.
We have shown that by imposing a Zeeman field to lift
the spin degeneracy along specific lines in momentum space,
the altermagnetic spin-splitting feature can be clearly
revealed by the frequency transitions in the quantum oscillations, regardless of the
SOC effects. To achieve this, the Landau level spacing $(eB/\hbar)\sqrt{4a^2-J^2}$ of altermagnets
should be smaller than the typical Zeeman splitting $\Delta$ that induces the Lifshitz
transition of the Fermi surface. This requires either a high-quality
sample that ensures high oscillating resolution for a small magnetic field $B$
or a relatively large Zeeman splitting. The latter can be
implemented by selecting materials with a large Land\'{e} $g$ factor
or introducing an exchange field via magnetic dopants~\cite{yu2010quantized}.
From the inter-orbit tunneling probability $Z=\text{exp}(-\frac{\pi}{B}\frac{\Delta^2}{2aJk_0^2})$,
a small size $k_0$ of the Fermi surface facilitates an efficient adjustment
of the frequency transition between two tunneling limits.
Although our work focuses on $d$-wave altermagnets,
the main conclusions can be applied to other symmetries such as
$g$- and $i$-wave symmetries, due to the universality
of the scenario. Moreover, we emphasize that the momentum-dependent spin splitting in altermagnetic materials is
on the eV scale, which is an order of magnitude larger than
the splitting induced by SOC~\cite{PhysRevX.12.031042,Manchon2015}. This implies that the frequency splitting
in altermagnetic materials is significantly more substantial.

\section*{Acknowledgement}
W. C. acknowledges financial support from
the National Natural Science Foundation of
China (No. 12222406 and No. 12074172), the Natural Science Foundation of Jiangsu Province (No. BK20233001),
the Fundamental Research Funds for the Central Universities (No. 2024300415), and
the National Key Projects for Research and Development of China (No. 2022YFA120470).  X. W.
acknowledges  financial  support  from  the  National  Natural
Science Foundation of China (No. 12188101).

\setcounter{equation}{0}
\renewcommand{\theequation}{\thesection \arabic{equation}}

\begin{appendix}
\section{Magnetic breakdown in altermagnets}\label{A}
In this section, we solve the problem of magnetic breakdown in altermagnets.
The Hamiltonian
$
\mathcal{{H}}(\bm{k})=ak_0(2k_y-k_0)+ Jk_0k_x \sigma_z+\Delta\sigma_x,
$ \emph{i.e.},
Eq.~(\ref{eq4}) of the main text,
yields hyperbolic equienergy contours given by
\begin{equation}
\frac{(k_y-k_0)^2}{(\Delta/2ak_0)^2}-\frac{k_x^2}{(\Delta/Jk_0)^2}=1.
\end{equation}
In this region, the Schr\"{o}dinger equation $\mathcal{H}(\bm{k}+\bm{A})\psi=E\psi$ (set $e=1, \hbar=1$)
including the effect of $B$ in the momentum representation reads
\begin{equation}\label{eqs2}
\begin{aligned}
iU_0\frac{\partial\psi}{\partial k_x}=U(k_x)\psi, \quad
U_0=\left(\begin{array}{cccccc}
2ak_0B        & 0  \\
0 & 2ak_0B      \\
\end{array}\right),\\
U(k_x)=E-ak_0(2k_y-k_0)- Jk_0k_x \sigma_z-\Delta \sigma_x.
\end{aligned}
\end{equation}
We define $\mathscr{H}(k_x)=U_0^{-1}U(k_x)$, Eq.~\eqref{eqs2} becomes
\begin{equation}
i\frac{\partial\psi}{\partial k_x}=\mathscr{H}(k_x)\psi,
\end{equation}
We further interpret $k_x$ as time $t$~\cite{PhysRevLett.116.236401}, the above equation
becomes the time-dependent Schr\"{o}dinger equation for the Hamiltonian $\mathscr{H}(t)$, where
\begin{equation}
\mathscr{H}(t)=\frac{1}{2ak_0B}[E-ak_0(2k_y-k_0)-\Delta \sigma_x-Jk_0t \sigma_z].
\end{equation}
Ignoring the terms proportional to the identity matrix $\sigma_0$ which only
change the energy by a constant, $\mathscr{H}(t)$ reduces to
\begin{equation}
\mathscr{H}(t)=-\frac{vt}{2}\sigma_z-\frac{\Delta'}{2}\sigma_x,
\end{equation}
with $v=\frac{J}{aB}$, $\Delta'=\frac{\Delta}{ak_0B}$. Next, we follow the procedure in Ref.~\cite{Shevchenko2009LandauZenerStckelbergI}.
Considering a general two-component wave function $\psi=(a_1, a_2)^T$, the Schr\"{o}dinger equation becomes
\begin{equation}
\left\{
\begin{array}{ll}
i\partial_t a_1=-\frac{vt}{2}a_1-\frac{\Delta'}{2}a_2,\\
i\partial_t a_2=-\frac{\Delta'}{2}a_1+\frac{vt}{2}a_2.
\end{array}
\right.
\end{equation}
Each component satisfies the second-order Weber equation as
\begin{equation}
\frac{d^2 a_{1,2}}{dz^2}+(2\delta\mp i+z^2)a_{1,2}=0,
\end{equation}
where
\begin{equation}\label{eqs8}
z=\sqrt{\frac{v}{2}}t, \quad \delta=\frac{\Delta'^2}{4v}.
\end{equation}
The solutions are combinations of parabolic cylinder functions~\cite{2007Table}. In regions far from
the transition point (where $z = 0$), $a_{1,2}$ take the following asymptotic expressions as
\begin{equation}\label{eqs9}
\begin{aligned}
a_1(-z_a)&\simeq A_+\Xi_1 e^{i\Phi(z_a)},\\
a_2(-z_a)&\simeq (-e^{-\frac{\pi}{2}\delta}A_+ +e^{\frac{\pi}{2}\delta}A_-)\Xi_2 e^{-i\Phi(z_a)},\\
a_1(z_a)&\simeq A_-\Xi_1 e^{i\Phi(z_a)},\\
a_2(z_a)&\simeq (-e^{\frac{\pi}{2}\delta}A_+ +e^{-\frac{\pi}{2}\delta}A_-)\Xi_2 e^{-i\Phi(z_a)},
\end{aligned}
\end{equation}
where $A_+$ and $A_-$ are coefficients, $z_a\gg1 $ and
\begin{equation}\label{eqs10}
\begin{aligned}
\Xi_1=\frac{\sqrt{2\pi}}{\Gamma(1+i\delta)}\text{exp}&(-\frac{\pi}{4}\delta),\quad \Xi_2=\frac{1}{\sqrt{\delta}}\text{exp}(-i\frac{\pi}{4}-\frac{\pi}{4}\delta),\\
\Phi(z_a)&=\frac{z_a^2}{2}+\delta {\rm ln} (\sqrt{2} z_a),
\end{aligned}
\end{equation}
with $\Gamma$ the gamma function. Meanwhile, the two instantaneous eigenstates $\varphi_1$ and $\varphi_2$ and
the corresponding eigenenergies of $\mathscr{H}(t)$ solved by $\mathscr{H}(t)\varphi_{1,2}=E\varphi_{1,2}$ are
\begin{equation}\label{s11}
\begin{aligned}
\varphi_1= \left(\begin{array}{cccccc}
 \sqrt{\frac{\Omega(t)+vt}{2\Omega(t)}} \\
 \sqrt{\frac{\Omega(t)-vt}{2\Omega(t)}} \\
\end{array}\right),\quad
\varphi_2= \left(\begin{array}{cccccc}
 \sqrt{\frac{\Omega(t)-vt}{2\Omega(t)}} \\
- \sqrt{\frac{\Omega(t)+vt}{2\Omega(t)}} \\
\end{array}\right),\\
E_{2,1}=\pm\frac{\Omega(t)}{2}, \quad \Omega(t)=\sqrt{\Delta'^2+v^2t^2}.
\end{aligned}
\end{equation}

Here, we are interested in the scattering between different orbits,
which is associated with the amplitudes denoted by $b_{1,2}$ under the basis of the instantaneous eigenstates in Eq.~\eqref{s11}.
A general time-dependent state can be expressed as $\psi(t)=b_1(t)\varphi_1(t)+b_2(t)\varphi_2(t)$. The adiabatic state evolution
from $t=t_i$ to $t=t_f$ can be described by the evolution matrix $U$, defined as follows
\begin{equation}\label{eqs12}
\begin{aligned}
&\left(\begin{array}{cccccc}
b_2(t_f) \\
b_1(t_f) \\
\end{array}\right)
=U(t_f,t_i)\left(\begin{array}{cccccc}
b_2(t_i) \\
b_1(t_i) \\
\end{array}\right),\\
U(t_f,t_i)&=\left(\begin{array}{cccccc}
e^{-i\int_{t_i}^{t_f}E_2(t)dt}  & 0 \\
0 & e^{-i\int_{t_i}^{t_f}E_1(t)dt}\\
\end{array}\right)\\
&=\left(\begin{array}{cccccc}
e^{-i\int_{t_i}^{t_f}\frac{\Omega(t)}{2}dt}  & 0 \\
0 & e^{i\int_{t_i}^{t_f}\frac{\Omega(t)}{2}dt}\\
\end{array}\right)
=e^{-i\int_{t_i}^{t_f}\frac{\Omega(t)}{2}}\sigma_z.
\end{aligned}
\end{equation}
For $t\rightarrow+\infty$ and $t\rightarrow-\infty$, the instantaneous eigenstates of Eq.~\eqref{s11} become
\begin{equation}
\begin{aligned}
\varphi_1&\simeq \left(\begin{array}{cccccc}
0 \\
1 \\
\end{array}\right),\quad
\varphi_2\simeq \left(\begin{array}{cccccc}
1 \\
0 \\
\end{array}\right), \quad \text{if} \quad t\rightarrow-\infty, \\
\varphi_1&\simeq \left(\begin{array}{cccccc}
1 \\
0 \\
\end{array}\right),\quad
\varphi_2\simeq \left(\begin{array}{cccccc}
0 \\
-1 \\
\end{array}\right), \quad \text{if} \quad t\rightarrow+\infty.
\end{aligned}
\end{equation}
The corresponding $\psi(t)$ are
\begin{equation}\label{s14}
\begin{aligned}
\psi(t)&= \left(\begin{array}{cccccc}
b_2 \\
b_1 \\
\end{array}\right), \quad \text{if} \quad t\rightarrow-\infty, \\
\psi(t)&= \left(\begin{array}{cccccc}
b_1 \\
-b_2 \\
\end{array}\right), \quad \text{if} \quad t\rightarrow+\infty.
\end{aligned}
\end{equation}
In the $t\rightarrow -\infty$ and $t\rightarrow +\infty$ limits,
comparing Eq.~\eqref{s14} with the eigenstate $\psi=(a_1, a_2)^T$
written in the Pauli representation yields
\begin{equation}\label{eqs15}
{\small
\begin{aligned}
\left(\begin{array}{cccccc}
b_2 (-z_a)\\
b_1 (-z_a)\\
\end{array}\right)
= \left(\begin{array}{cccccc}
a_1 (-z_a)\\
a_2 (-z_a)\\
\end{array}\right), \quad
\left(\begin{array}{cccccc}
b_2 (z_a)\\
b_1 (z_a)\\
\end{array}\right)
= \left(\begin{array}{cccccc}
-a_2 (z_a)\\
a_1 (z_a)\\
\end{array}\right),
\end{aligned}
}
\end{equation}
where the notation $z_a$ is defined in Eq.~\eqref{eqs8} and Eq.~\eqref{eqs9}.

The inter-orbit scattering occurs only in the four regions near the Lifshitz transition points [dashed
boxes in Fig.~1 of the main text],
which can be captured by a scattering matrix $\mathcal{S}$ connecting the incoming ($t=0^-$)
and outgoing ($t=0^+$) states as (note that the scattering time is negligible compared to the cyclotron period)
\begin{equation}
\begin{aligned}
\left(\begin{array}{cccccc}
b_2 (0^+)\\
b_1 (0^+)\\
\end{array}\right)
= \mathcal{S}\left(\begin{array}{cccccc}
b_2 (0^-)\\
b_1 (0^-)\\
\end{array}\right).
\end{aligned}
\end{equation}
Consider a wave function evolving from $t=-t_a$ to $t_a$, which contains a single scattering event
that is assumed to be around $t=0$. Away from the vicinity of $t=0$, the inter-orbit tunneling
can be neglected and the evolution of the wave function is adiabatic.
Therefore, the state evolution can be divided into three steps,
the adiabatic evolutions from $t=-t_a$ to $0^-$ and from $t=0^+$ to $t_a$ connected by
the scattering from $t=0^-$ to $0^+$. Expressing it in a compact form yields
\begin{equation}\label{eqs17}
\begin{aligned}
\left(\begin{array}{cccccc}
b_2 (t_a)\\
b_1 (t_a)\\
\end{array}\right)
=U(t_a, 0^+)\mathcal{S}U(0^-, -t_a) \left(\begin{array}{cccccc}
b_2 (-t_a)\\
b_1 (-t_a)\\
\end{array}\right).
\end{aligned}
\end{equation}
According to Eq.~\eqref{eqs12}, the adiabatic time evolution is described by the unitary matrix $U$ as
\begin{equation}
\begin{aligned}
U(0_-,-t_a)&=U(t_a,0^+)=
e^{-i\int_{0}^{t_a}\frac{\Omega(t)}{2}}\sigma_z=e^{-i\zeta(t_a)}\sigma_z,\\
\zeta(t_a)&=\frac{z_a^2}{2}+\delta {\rm ln} (\sqrt{2} z_a)-\frac{1}{2}\delta(\rm{ln}\delta-1)
\end{aligned}
\end{equation}
where we have replaced $t_a$ with $z_a=\sqrt{\frac{v}{2}}t_a$ [see also Eq.~\eqref{eqs8} and Eq.~\eqref{eqs9}].
Using the asymptotic equality~\eqref{eqs15}, Eq.~\eqref{eqs17} becomes
\begin{equation}
\begin{aligned}
&\left(\begin{array}{cccccc}
-a_2 (z_a)\\
a_1 (z_a)\\
\end{array}\right)\\
=\left(\begin{array}{cccccc}
e^{-i\zeta(t_a)}  & 0 \\
0 & e^{i\zeta(t_a)}\\
\end{array}\right)&\mathcal{S}
\left(\begin{array}{cccccc}
e^{-i\zeta(t_a)}  & 0 \\
0 & e^{i\zeta(t_a)}\\
\end{array}\right) \left(\begin{array}{cccccc}
a_1 (-z_a)\\
a_2 (-z_a)\\
\end{array}\right).
\end{aligned}
\end{equation}
Comparing this equality to Eq.~\eqref{eqs9}, we obtain the scattering matrix $\mathcal{S}$~\cite{Shevchenko2009LandauZenerStckelbergI} as
\begin{equation}\label{eqs20}
\begin{aligned}
\mathcal{S}=\left(\begin{array}{cccccc}
\sqrt{1-Z}e^{-i\omega}  & -\sqrt{Z} \\
\sqrt{Z} & \sqrt{1-Z}e^{i\omega}\\
\end{array}\right),\\
 \omega=\frac{\pi}{4}+\delta({\rm ln}\delta-1)+{\rm arg}[\Gamma(1-i\delta)]
\end{aligned}
\end{equation}
which is Eq.~(4) in the main text.

\setcounter{equation}{0}
\renewcommand{\theequation}{\thesection \arabic{equation}}
\section{General quantization condition}\label{B}
By inserting Eq.~\eqref{eqs20} into Eq.~(\ref{eq8}) in the main text, we obtain
\begin{widetext}
\begin{equation}
\begin{aligned}
1+e^{i(\Theta_1+\Theta_2+\Theta_3+\Theta_4+\Theta_5+\Theta_6+\Theta_7+\Theta_8)}&-(1-Z)^2(e^{i(4\omega+\Theta_1+\Theta_3+\Theta_5+\Theta_7)}+e^{i(-4\omega+\Theta_2+\Theta_4+\Theta_6+\Theta_8)})\\
+Z(1-Z)e^{2i\omega}(e^{i(\Theta_2+\Theta_3+\Theta_5+\Theta_7)}&+e^{i(\Theta_1+\Theta_4+\Theta_5+\Theta_7)}+e^{i(\Theta_1+\Theta_3+\Theta_6+\Theta_7)}+e^{i(\Theta_1+\Theta_3+\Theta_5+\Theta_8)})\\
+Z(1-Z)e^{-2i\omega}(e^{i(\Theta_2+\Theta_4+\Theta_6+\Theta_7)}&+e^{i(\Theta_2+\Theta_4+\Theta_5+\Theta_8)}+e^{i(\Theta_2+\Theta_3+\Theta_6+\Theta_8)}+e^{i(\Theta_1+\Theta_4+\Theta_6+\Theta_8)})\\
+Z(1-Z)(e^{i(\Theta_2+\Theta_4+\Theta_5+\Theta_7)}&+e^{i(\Theta_1+\Theta_3+\Theta_6+\Theta_8)}+e^{i(\Theta_1+\Theta_4+\Theta_6+\Theta_7)}+e^{i(\Theta_2+\Theta_3+\Theta_5+\Theta_8)})\\
+Z^2(e^{i(\Theta_2+\Theta_3+\Theta_6+\Theta_7)}&+e^{i(\Theta_1+\Theta_4+\Theta_5+\Theta_8)})=0,
\end{aligned}
\end{equation}
\end{widetext}
and then
\begin{widetext}
{\small
\begin{equation}\label{eqs23}
\begin{aligned}
\cos\left(\frac{\Theta_1+\Theta_2+\Theta_3+\Theta_4+\Theta_5+\Theta_6+\Theta_7+\Theta_8}{2}\right)-(1-Z)^2\cos\left(\frac{\Theta_1+\Theta_3+\Theta_5+\Theta_7-\Theta_2-\Theta_4-\Theta_6-\Theta_8}{2}+4\omega\right)\\
+Z(1-Z)\left[\cos\left(\frac{\Theta_2+\Theta_3+\Theta_5+\Theta_7-\Theta_1-\Theta_4-\Theta_6-\Theta_8}{2}+2\omega\right)+\cos\left(\frac{\Theta_1+\Theta_4+\Theta_5+\Theta_7-\Theta_2-\Theta_3-\Theta_6-\Theta_8}{2}+2\omega\right) \right.\\
\left.+\cos\left(\frac{\Theta_1+\Theta_3+\Theta_6+\Theta_7-\Theta_2-\Theta_4-\Theta_5-\Theta_8}{2}+2\omega\right)+\cos\left(\frac{\Theta_1+\Theta_3+\Theta_5+\Theta_8-\Theta_2-\Theta_4-\Theta_6-\Theta_7}{2}+2\omega\right)
\right]\\
+Z(1-Z)\left[\cos\left(\frac{\Theta_2+\Theta_4+\Theta_5+\Theta_7-\Theta_1-\Theta_3-\Theta_6-\Theta_8}{2}\right)+\cos\left(\frac{\Theta_1+\Theta_4+\Theta_6+\Theta_7-\Theta_2-\Theta_3-\Theta_5-\Theta_8}{2}\right)\right]\\
-Z^2\cos\left(\frac{\Theta_2+\Theta_3+\Theta_6+\Theta_7-\Theta_1-\Theta_4-\Theta_5-\Theta_8}{2}\right)=0.       \quad\quad\quad\quad\quad\quad\quad\quad\quad\quad\quad\quad\quad\quad\quad
\end{aligned}
\end{equation}
}
\end{widetext}
\normalsize{
For a closed trajectory $O$ in the momentum space, the accumulated phase $\Theta_O$ can be expressed as~\cite{PhysRevB.97.144422}
\begin{equation}
\Theta_O=l^2S(O)-2\pi\gamma,
\end{equation}
where $S(O)$ is the area encircled by the trajectory $O$.
We write all 16 possible trajectories explicitly as

\allowdisplaybreaks
\begin{align}\label{eqs25}
\Theta_1+\Theta_3+\Theta_5+\Theta_7=l^2S_5-2\pi\gamma, \notag \\  \Theta_2+\Theta_4+\Theta_6+\Theta_8=l^2(S_5+4S_0)-2\pi\gamma,\notag \\
\Theta_2+\Theta_3+\Theta_5+\Theta_7=l^2(S_5+S_0)-2\pi\gamma, \notag \\ \Theta_1+\Theta_4+\Theta_6+\Theta_8=l^2(S_5+3S_0)-2\pi\gamma, \notag \\
\Theta_1+\Theta_4+\Theta_5+\Theta_7=l^2(S_5+S_0)-2\pi\gamma, \notag\\  \Theta_2+\Theta_3+\Theta_6+\Theta_8=l^2(S_5+3S_0)-2\pi\gamma, \notag\\
\Theta_1+\Theta_3+\Theta_6+\Theta_7=l^2(S_5+S_0)-2\pi\gamma, \notag\\  \Theta_2+\Theta_4+\Theta_5+\Theta_8=l^2(S_5+3S_0)-2\pi\gamma, \notag\\
\Theta_1+\Theta_3+\Theta_5+\Theta_8=l^2(S_5+S_0)-2\pi\gamma, \notag\\  \Theta_2+\Theta_4+\Theta_6+\Theta_7=l^2(S_5+3S_0)-2\pi\gamma, \notag\\
\Theta_2+\Theta_4+\Theta_5+\Theta_7=l^2(S_5+2S_0)-2\pi\gamma, \notag\\ \Theta_1+\Theta_3+\Theta_6+\Theta_8=l^2(S_5+2S_0)-2\pi\gamma, \notag\\
\Theta_1+\Theta_4+\Theta_6+\Theta_7=l^2(S_5+2S_0)-2\pi\gamma, \notag\\  \Theta_2+\Theta_3+\Theta_5+\Theta_8=l^2(S_5+2S_0)-2\pi\gamma, \notag\\
\Theta_2+\Theta_3+\Theta_6+\Theta_7=l^2(S_5+2S_0)-2\pi\gamma, \notag\\  \Theta_1+\Theta_4+\Theta_5+\Theta_8=l^2(S_5+2S_0)-2\pi\gamma.
\end{align}
Substituting Eq.~\eqref{eqs25} to Eq.~\eqref{eqs23}, we obtain the quantization condition
\begin{equation}
\begin{aligned}
\cos[l^2(S_5+2S_0)-2\pi\gamma]-(1-Z)^2\cos(2l^2S_0-4\omega)\\
+4Z(1-Z)\cos(l^2S_0-2\omega)+Z(2-3Z)=0,
\end{aligned}
\end{equation}
which is Eq.~(\ref{eq10}) in the main text.
}

\setcounter{equation}{0}
\renewcommand{\theequation}{\thesection \arabic{equation}}
\section{The calculation of Berry phase}\label{C}
Consider the whole Hamiltonian $H(\bm{k})+H_R(\bm{k})$ with the second term
corresponding to the Rashba SOC as
$
H_R(\bm{k})=\lambda(k_y\sigma_x-k_x\sigma_y).
$
The eigenenergies are $\mathcal{E}_{\pm}=a(k_x^2+k_y^2)\pm\sqrt{(\Delta+\lambda k_y)^2+(\lambda k_x)^2+(Jk_xk_y)^2}$.
The Berry connection $\boldsymbol{A}_n(\boldsymbol{k})$ of an eigenstate $\psi_n$ is defined
as $\boldsymbol{A}_n(\boldsymbol{k})=i\langle\psi_n|\nabla_{\boldsymbol k}|\psi_n\rangle$
and the corresponding Berry curvature $\boldsymbol{\Omega}_n(\boldsymbol{k})$ is defined as $\boldsymbol{\Omega}_n(\boldsymbol{k})=\nabla\times\boldsymbol{A}_n(\boldsymbol{k})$.
Without the in-plane Zeeman filed with $\Delta=0$, $\boldsymbol{\Omega}(\boldsymbol{k})$ reads~\cite{PhysRevX.12.040501}
\begin{equation}\label{eqs52}
\boldsymbol{\Omega}(\boldsymbol{k})_{\pm}=\mp\frac{J\lambda^2k_xk_y}{2[J^2k_x^2k_y^2+\lambda^2(k_x^2+k_y^2)]^{\frac{3}{2}}},
\end{equation}
which diverges at the origin. To avoid the singularity, we introduce an infinitely
small term $\eta\sigma_z$ ($\eta\rightarrow 0^+$) to the Hamiltonian and the Berry curvature becomes
\begin{equation}\label{eqs53}
\boldsymbol{\Omega}(\boldsymbol{k})_{\pm}=\mp\frac{\lambda^2 (J k_xk_y+\eta)}{2[(J k_xk_y+\eta)^2+\lambda^2(k_x^2+k_y^2)]^{\frac{3}{2}}}.
\end{equation}
With the analytic Berry curvature, the Berry phase $\phi_B$ can be obtained by the integral as
\begin{equation}\label{eqs54}
\phi_{B\pm}=\lim_{\eta\rightarrow0^+}\int\int_{S_L} dk_xdk_y\boldsymbol{\Omega}(\boldsymbol{k})_{\pm}=\mp\pi,
\end{equation}
where $S_L$ is the area encircled by the cyclotron orbits $L=\alpha_{1,2}$ at the Fermi energy in Fig.~\ref{fig3}(a) of the main text.

For a finite $\Delta$, the Berry curvature becomes
\begin{equation}\label{eqs55}
\boldsymbol{\Omega}(\boldsymbol{k})_{\pm}=\mp\frac{J\lambda k_x(\lambda k_y+\Delta)}{2[J^2k_x^2k_y^2+\lambda^2k_x^2+(\lambda k_y+\Delta)^2]^{\frac{3}{2}}},
\end{equation}
which shifts the singular point of $\boldsymbol{\Omega}(\boldsymbol{k})$ to ($0, -\Delta/\lambda$).
As long as $\Delta<\lambda k_0$,
the Berry phase $\phi_B$ remains the same as that for $\Delta=0$.
%At the critical point with $\Delta=\lambda k_0$, The orbits of $E_+$ and $E_-$ merge
%and the Berry phase of the merged orbit is $\phi_B=\phi_{B+}+\phi_{B-}=0$.
For $\Delta>\lambda k_0$, the singular point is outside of the orbit $L=\zeta_{1,2}$ in Fig.~\ref{fig3}(c) of the main text.
Therefore, the Berry phase $\phi_B$ is
\begin{equation}\label{eqs56}
\phi_{B\pm}=\int\int_{S_L} dk_xdk_y\boldsymbol{\Omega}(\boldsymbol{k})_{\pm}=0.
\end{equation}\par
%\bibliography{ref}

%

\end{appendix}

\end{document}